\documentclass[10pt,twocolumn]{article} 
\usepackage{amsmath,amssymb}
\usepackage[utf8]{inputenc}
\usepackage{authblk}

\usepackage{blindtext}
\usepackage{abstract}
\usepackage{geometry}
\usepackage[colorlinks,allcolors=black,bookmarks=false,hypertexnames=true]{hyperref} 

\usepackage[backend=biber, style=phys, sorting=none]{biblatex}
\usepackage{textcomp}
\usepackage{tabularx}
\usepackage{url}
\usepackage{graphicx}
\usepackage{layout}
\usepackage{blindtext}
\usepackage[margin=10pt,font=footnotesize,labelfont=bf,format=plain]{caption} 

\let\svthefootnote\thefootnote
\newcommand\freefootnote[1]{%
  \let\thefootnote\relax%
  \footnotetext{#1}%
  \let\thefootnote\svthefootnote%
}

\title{A quantitative comparison of time-of-flight momentum microscopes and hemispherical analyzers for time- and angle-resolved photoemission spectroscopy experiments}

\author{J. Maklar}
\author{S. Dong}
\author{S. Beaulieu}
\author{T. Pincelli}
\author{M. Dendzik$^*$}
\author{Y.W. Windsor}
\author{R.P. Xian$^\dagger$}
\author{M. Wolf}
\author{R. Ernstorfer}
\author{L. Rettig}
\affil{Fritz-Haber-Institut der Max-Planck-Gesellschaft, Faradayweg 4-6, D-14195 Berlin, Germany}

\date{\today}
\begin{document}

\twocolumn[
\maketitle
\begin{@twocolumnfalse}
    \maketitle
    \begin{abstract}

Time-of-flight-based momentum microscopy has a growing presence in photoemission studies, as it enables parallel energy- and momentum-resolved acquisition of the full photoelectron distribution. Here, we report table-top extreme ultraviolet (XUV) time- and angle-resolved photoemission spectroscopy (trARPES) featuring both a hemispherical analyzer and a momentum microscope within the same setup. We present a systematic comparison of the two detection schemes and quantify experimentally relevant parameters, including pump- and probe-induced space-charge effects, detection efficiency, photoelectron count rates, and depth of focus. We highlight the advantages and limitations of both instruments based on exemplary trARPES measurements of bulk WSe$_2$. Our analysis demonstrates the complementary nature of the two spectrometers for time-resolved ARPES experiments. Their combination in a single experimental apparatus allows us to address a broad range of scientific questions with trARPES.
\end{abstract}
\end{@twocolumnfalse}
]

\freefootnote{Current address: \\$^*$Department of Applied Physics, KTH Royal Institute of Technology, SE-16440, Stockholm, Kista, Sweden\\
$^\dagger$Department of Neurobiology, Northwestern University, IL-60208 Evanston, USA}

\section{Introduction} \label{sec:introduction}
Angle-resolved photoemission spectroscopy (ARPES) is a key technique to investigate the electronic structure of solids. By extracting the kinetic energy and angular distribution of emitted photoelectrons, one gains direct access to the spectral function and, in particular, the quasiparticle band structure\cite{damascelli2004}. Combining this technique with a pump-probe approach allows studying the electron dynamics after optical excitation on a femtosecond timescale. In recent years, time-resolved ARPES (trARPES) has been successfully applied to many fields in materials science, such as control of quantum matter\cite{ ulstrup2016ultrafast, mahmood2016, bertoni2016,Reimann2018, beaulieu2020},  photo-induced phase transitions\cite{schmitt2008, rohwer2011, cortes2011, smallwood2012, mathias2016self, nicholson2018, tengdin2018critical} and the investigation of electronic states and phases not accessible in equilibrium\cite{sobota2012, gierz2013, ligges2018, shi2019}. Advances in laser-based extreme ultraviolet (XUV) sources using high harmonic generation in noble gases\cite{mcpherson1987studies,ferray1988multiple,haight1994tunable} now enable space-charge free photoemission up to MHz repetition rates at high time and energy resolution (10s of fs/meV) and at photon energies up to the far XUV\cite{eich2014time, rohde2016, corder2018, mills2019, buss2019, cucini2019, puppin2019, sie2019, keunecke2020, peli2020time}.

\begin{figure*}[h!]
\centering
\includegraphics[width=1\textwidth]{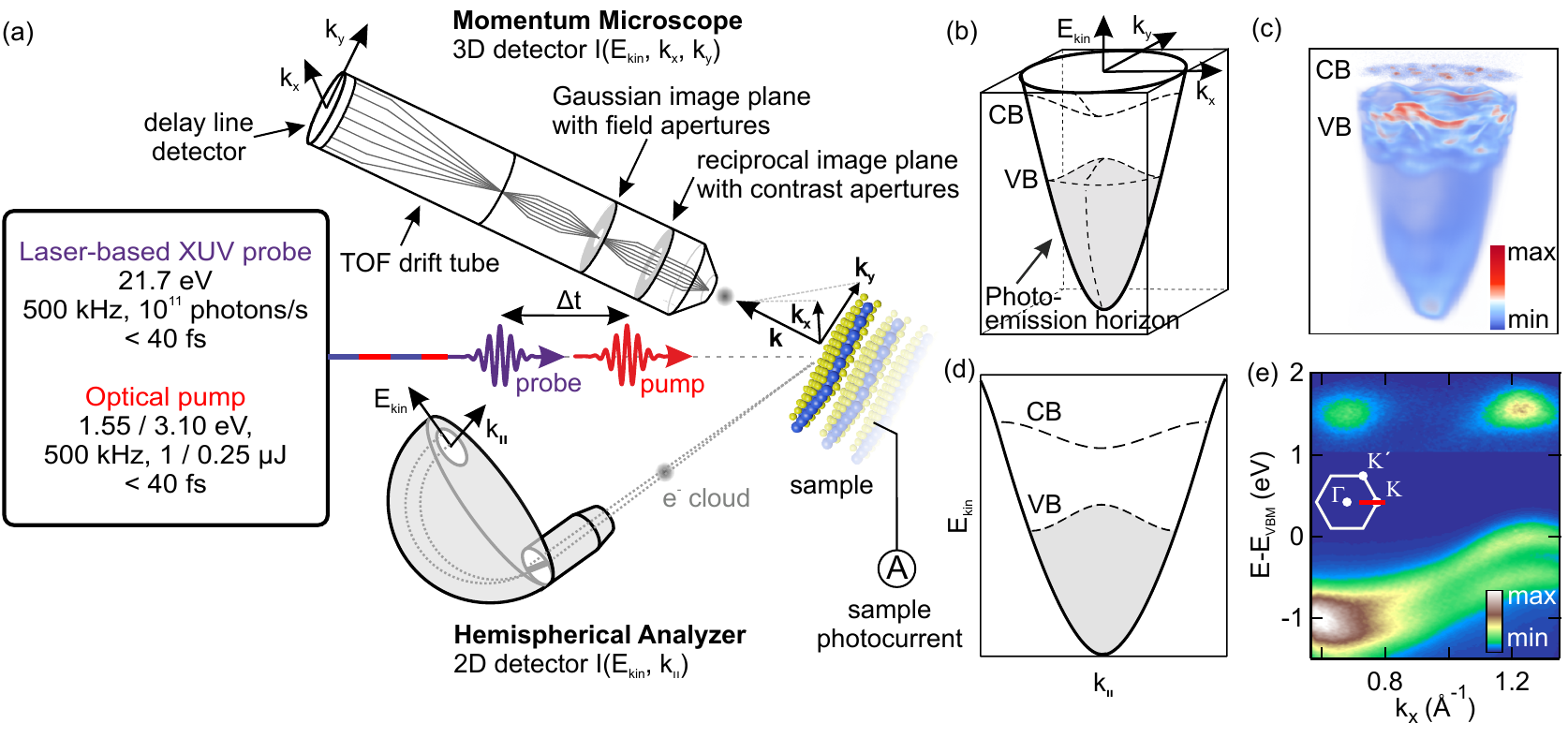}
\caption{(a) Schematic layout of the setup. Note that in the real experiment the angle of incidence for measurements with the MM is fixed at 65°. (b) Illustration and (c) experimental data of a 3D dataset of WSe$_2$ acquired with the MM. (d) Sketch and (e) data of a 2D energy-momentum cut acquired with the HA. The momentum direction within the hexagonal BZ of WSe$_2$ is indicated in red. The excited-state signal above the valence band maximum of the exemplary datasets (pump-probe delay $t=0\,$fs, absorbed fluence $F_{\mathrm{abs}}=150$\,\textmu J$/$cm$^{2}$) is enhanced by a factor of (c) 100 and (e) 75. In all MM measurements, the extractor voltage is $V_{\mathrm{extr}}=6$\,kV and the sample-extractor distance is 4\,mm with the sample surface aligned perpendicular to the optical axis of the instrument.}
\label{fig:setup}
\end{figure*}

The most commonly used electron spectrometer in trARPES is the hemispherical analyzer (HA)\cite{hufner2013photoelectron}. Here, the photoelectrons enter an electrostatic lens system followed by two hemispherical deflector electrodes acting as a dispersive band-pass energy filter, as sketched in Fig.\,\ref{fig:setup}(a). Subsequently, the electrons are projected onto a 2D multi-channel plate (MCP) detector, which allows parallel detection of kinetic energy and emission angle. This detection scheme is rather inefficient, as only a single two-dimensional (2D) cut in a narrow energy and momentum window of the 3D photoelectron distribution can be simultaneously captured. 

The more recent detection scheme based on a time-of-flight (ToF) energy analyzer overcomes this limitation, however, requires a pulsed light source with an appropriate repetition rate\cite{tusche2016multi}: The momentum microscope (MM) is based on a cathode-lens electron microscope\cite{kotsugi2003microspectroscopic, kromker2008development, tusche2015spin, schonhense2015}. By applying a high positive voltage to an electrostatic objective lens placed close to the sample surface, all emitted photoelectrons are steered into the lens system resulting in an acceptance of the complete $2\pi$ solid angle. In analogy to optical microscopy, a reciprocal image is generated in the back focal plane of the objective lens, corresponding to the surface-projected band structure. Next, the photoelectrons pass through a field-free ToF drift tube. Finally, their 2D momentum distribution and kinetic energy (encoded in the arrival time) are detected at a single-electron level using an MCP stack combined with a position-sensitive delay-line detector (DLD). Ultimately, the ToF-MM enables parallel acquisition of the 3D photoelectron distribution $I\left(E_{\mathrm{kin}},k_x,k_y\right)$ across the full accessible in-plane momentum range (at low kinetic energies limited by the parabola of the photoemission horizon) and within a large energy range from the threshold energy to the hard X-ray regime \cite{schonhense2015, medjanik2019progress, kutnyakhov2020}, as illustrated in Fig.\,\ref{fig:setup}(b). 

In principle, trARPES is expected to benefit greatly from the improved parallelization in data acquisition of the ToF-MM for several reasons: (i) The excited-state signal is usually orders of magnitude lower than that of the occupied states in equilibrium\cite{grubis2015, bertoni2016, ligges2018, nicholson2018, liu2019direct}, which necessitates efficient detection. (ii) Prediction of the relevant energy-momentum regions of photoexcited states can be difficult, and a time-resolved mapping of the entire first Brillouin zone (BZ) with a HA is typically not feasible. (iii) Various photoinduced electronic processes can occur simultaneously, spread over a large energy-momentum range, which are now accessible within a single measurement. However, while the MM theoretically constitutes the ultimate photoelectron detector, certain limitations, such as increased space-charge effects\cite{schonhense2018, kutnyakhov2020} and constraints of the DLD detection rate, compromise the experimental practicability in particular for pump-probe experiments. Therefore, a detailed benchmark of these two photoelectron detection schemes is of great interest.

In this article, we present a table-top XUV trARPES setup that combines a ToF-MM and a conventional HA and investigate their respective operational capabilities. We quantify critical parameters, such as depth of focus, experimental count rates, acquisition times, and space-charge effects. By two exemplary trARPES experiments -- excited-state band structure mapping at a fixed time delay and tracking of the excited population dynamics -- we demonstrate the advantages and limitations of both instruments and illustrate the benefits of combining both types of analyzers. After an overview of our experimental setup in Sec.\,\ref{sec:setup}, we will introduce some important aspects specific to the MM in Sec.\,\ref{sec:dof}. Section\,\ref{sec:comparison} finally compares the two spectrometers based on our experimental data, followed by a discussion in Sec.\,\ref{sec:discussion}.

\section{Description of the experimental setup} \label{sec:setup}

The table-top XUV light source consists of an optical parametric chirped pulse amplifier (OPCPA) generating fs light pulses at 1.55\,eV and 500\,kHz at an average power of 20\,W (40\,\textmu J pulse energy)\cite{puppin2015}. A beamsplitter at the exit of the OPCPA extracts a portion of the pulse energy as a 1.55\,eV or frequency-doubled 3.1\,eV synchronized optical pump. The probe pulses are frequency-doubled in a beta barium borate crystal and focused onto a high-pressure argon jet for up-conversion to the XUV via high harmonic generation. By a combination of a multilayer mirror and metallic (Sn) filters, only the 7th harmonic (21.7\,eV) is transmitted to the analysis chamber\cite{puppin2019}. Then, the pump and probe beams are focused onto the sample in a near-collinear geometry, and the emitted photoelectrons are detected with a HA (SPECS PHOIBOS 150 2D-CCD) or a ToF-MM (SPECS METIS 1000). The MM is mounted on a linear translation stage connected to the analysis chamber by a vacuum bellow and can be retracted to avoid collision with the cryogenic 6-axis carving manipulator when using the HA. 

For time-resolved studies with the HA, measurements are performed for a series of pump-probe delays. When using the MM, we continuously scan a defined pump-probe delay window, whereas the current delay is stored for each measurement event by an analog-to-digital conversion of the delay stage position. The detection unit of the MM features an MCP followed by a DLD. Each registered event directly corresponds to a single photoelectron. Saving this data stream at a single-event level permits event-wise correction and calibration, and selective binning later during analysis\cite{xian2019open, xian2019symmetry, kutnyakhov2020}. The operating principle of the DLD limits the count rate to a single electron per pulse\cite{sobottka1988delay, oelsner2001microspectroscopy}, resulting in maximum rates of $\sim5 \times 10^{5}$\,cts/s, corresponding to the repetition rate of the laser system. For the case of the HA, the photoelectrons are first multiplied in an MCP and subsequently accelerated onto a phosphor screen, which is imaged by a CCD camera. Thus, a single photoelectron generates several counts spread over adjacent pixels. To obtain an estimate of the actual photoelectron count rate, we calibrated the CCD response in the regime of distinct single-electron events. To quantify relevant experimental parameters of both spectrometers (see Sec.\,\ref{sec:comparison}), we introduce the metrics \textit{emitted electrons per pulse}  $e_{\mathrm{tot}}$, i.e., the total photoelectron yield per pulse obtained from the sample photocurrent, and \textit{detected electrons per pulse} $cts_{\mathrm{MM}}$ and $cts_{\mathrm{HA}}$, corresponding directly to the count rate of the MM and to the rescaled CCD count rate of the HA, respectively.

The material used for the benchmark study is bulk tungsten diselenide (2H-WSe$_2$). This layered semiconductor exhibits an indirect bandgap\cite{manzeli2017}, a sharp electronic band structure and a distinct electronic response upon near-infrared optical excitation\cite{bertoni2016}. Exemplary datasets acquired with both detectors on WSe$_2$ at temporal pump-probe overlap are shown in Figs.\,\ref{fig:setup}(c,e). The MM captures the entire photoemission horizon (momentum disk with radius $k_{\parallel, max}\approx 2.15\,$\AA$^{-1}$), exceeding the first BZ of WSe$_2$, and the full energy range from the pump-pulse-induced population in the conduction band (CB) to the secondary electron cutoff. In contrast, the HA covers an energy window of a few electron volts (at a reasonable energy resolution) and a narrow momentum range resulting from the limited acceptance angle of $\pm$\,15° (Wide Angle Mode). The momentum resolution orthogonal to the dispersing direction is determined by the width of the slit located at the entrance of the spherical deflector. All HA data were recorded with a slit width of 0.5\,mm, corresponding to a momentum integration of $\approx$ 0.04\,\AA$^{-1}$, and a pass energy of 30\,eV.

Using the MM, the angle between the pump and probe beams and the sample surface normal is fixed at 65°. For comparability between the detectors, we align the sample in a similar geometry in the HA measurements, which yields the $\Sigma$-K momentum cut shown in Fig.\,\ref{fig:setup}(e). All samples are cleaved at room temperature in ultra high vacuum ($<1$ $\times$ 10$^{-10}$\,mbar).

Whereas the energy resolution of our trARPES setup is limited by the bandwidth of the XUV probe pulses to $\sim\,$150\,meV, the HA offers an improved momentum resolution over the MM. Based on band structure data, we estimate an effective momentum resolution of the MM and the HA of 0.08\,\AA$^{-1}$ and 0.04\,\AA$^{-1}$ ($\sim$1\,°), respectively. The ultimate instrument resolution is reported as $<\,4\,\times\,10^{-3}$\,\AA$^{-1}$ ($<0.1$\,°) for the HA and $<\,5\,\times\,10^{-3}$\,\AA$^{-1}$ for the MM\cite{schonhense2015, tusche2015spin, tusche2019imaging}. However, achieving such optimal conditions with the MM requires very high extractor voltages and tedious optimization of the lens settings and corrector elements.

\section{Depth of focus in Momentum Microscopy} \label{sec:dof}

Before starting our systematic comparison of the two spectrometers, we introduce further features of the MM arising from the similarity to optical microscopy. Firstly, both a reciprocal and a Gaussian real-space image plane form consecutively in the electron-optical lens column, which can be selectively projected onto the DLD. Thus, by choice of lens settings, the instrument can be used either for band structure mapping or to investigate the real-space distribution of photoelectrons via photoemission electron microscopy (PEEM)\cite{Feng2019}. Secondly, apertures can be inserted in both image planes, which enables trARPES at high spatial selectivity and time- and momentum-resolved PEEM.

\begin{figure}[t!]
\centering
\includegraphics[width=1\columnwidth]{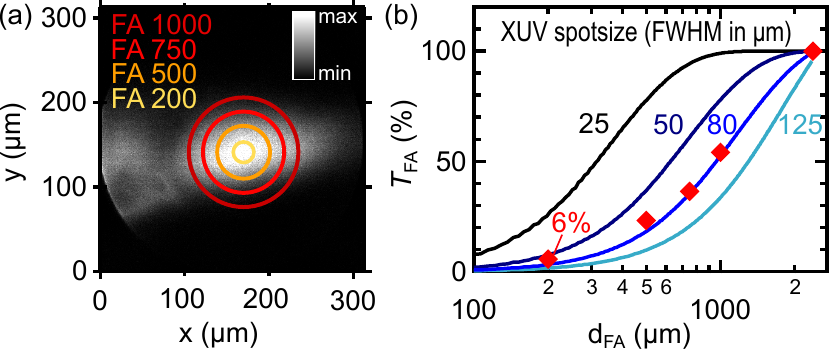}
\caption{(a) PEEM image of the elongated XUV beam footprint (focal diameter of $\approx 80$\,\textmu m) at an incidence angle of 65° on a WSe$_2$ sample at a magnification of 7.6. Projected field aperture sizes are illustrated in color (diameters in \textmu m). (b) Calculated transmission as a function of field aperture diameter for selected probe spot sizes, taking into account the angle of incidence. Experimentally determined values, corresponding to the apertures indicated in panel (a), are marked by red diamonds.}
\label{fig:FAs}
\end{figure}

We first focus on the use of field apertures inserted into the Gaussian image plane, which can be used to study the electronic band structure of spatially inhomogeneous or small samples below the size of the probe spot down to the micrometer range, see Fig.\,\ref{fig:FAs}(a). The electron transmission $T_{\mathrm{FA}}$ for various field apertures and for various probe spot sizes is shown in Fig.\,\ref{fig:FAs}(b).

\begin{figure}[t!]
\centering
\includegraphics[width=1\columnwidth]{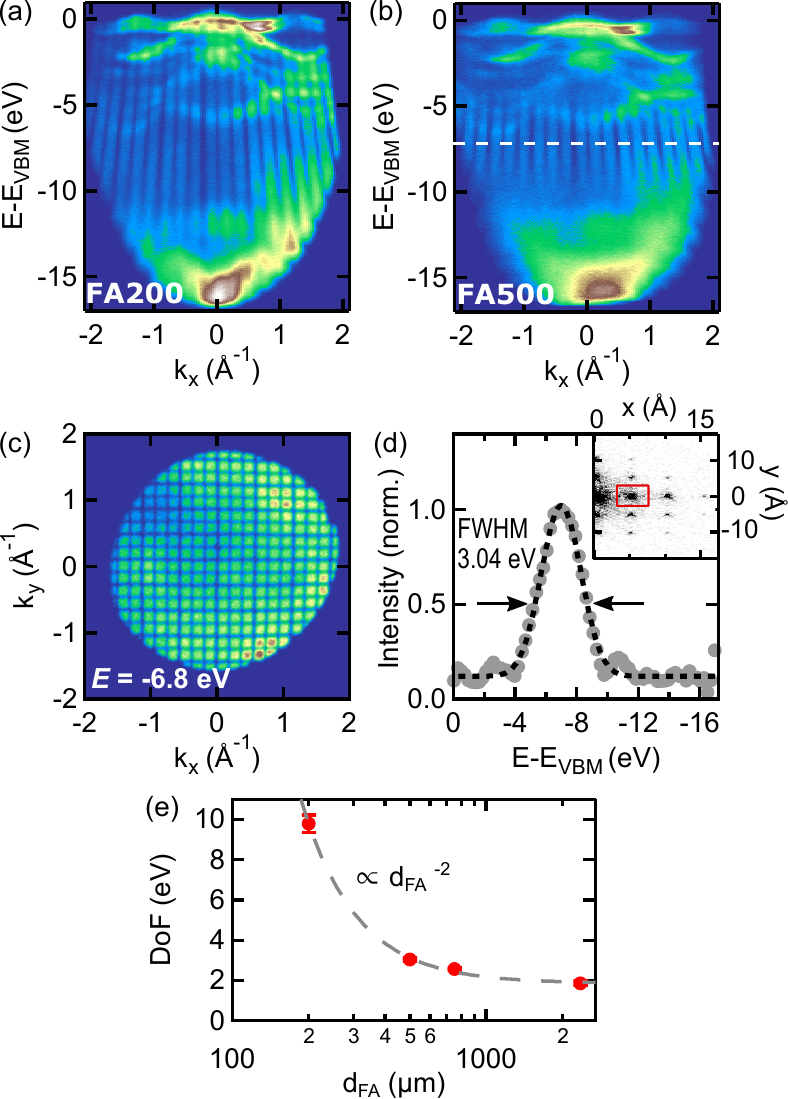}
\caption{2D cut of a MM measurement along the K-$\Gamma$-K direction with a square grid in the momentum image plane for field apertures of diameters (a) 200\,\textmu m and (b) 500\,\textmu m. (c) Iso-energy contour at the focus energy (sharpest momentum image), see the white dashed line in panel (b), for d$_{\mathrm{FA}}=500$\,\textmu m. (d) Intensity of the Fourier transform peak corresponding to the grid spacing, see the red box in the inset, as a function of energy. The FWHM of the peak is extracted from a Gaussian fit (black dashed curve). The inset shows the Fourier transform of the iso-energy contour in c. (e) Depth of focus (FWHM) versus aperture diameter with an inverse quadratic fit.}
\label{fig:DoF}
\end{figure}

The effective source size, defined by the field aperture or the spot size, also determines the depth of focus (DoF), i.e., the energy window with sharp momentum resolution, resulting from the chromatic aberrations of the electron lenses. To investigate the DoF, we insert a grid in the momentum image plane, and analyze the sharpness of the resulting grid lines as a function of kinetic energy for various field apertures, shown in Fig.\,\ref{fig:DoF}. For the aperture diameter d$_{\mathrm{FA}}=200$\,\textmu m, we observe sharp grid lines superimposed on the band structure of WSe$_2$ reaching from the valence band (VB) down to almost the entire secondary electron tail. However, with increasing aperture size, the energy window of sharp momentum imaging narrows. To quantify this trend, we perform a 2D Fourier transform of the iso-energy contours, and analyze the magnitude of the spatial frequency peaks corresponding to the grid periodicity as a function of energy, shown in Fig.\,\ref{fig:DoF}(c-d). Similar to the depth of field in optical imaging\cite{murphy2002fundamentals}, we find that the DoF follows an inverse square dependence of the aperture diameter, see Fig.\,\ref{fig:DoF}(e).

To achieve a uniform performance in a typical range of interest of few eV, it is necessary to have a DoF of $\sim 10$\,eV. For this, the effective source size has to be reduced to $\sim 25$\,\textmu m, which corresponds to a field aperture diameter of 200\,\textmu m for the chosen magnification settings. At the given spot size of $80\times 80$\,\textmu m$^2$, this reduces the photoelectron transmission to $T_{\mathrm{FA}}=6$\,\% of the total yield, as shown in Fig.\,\ref{fig:FAs}(b). However, to compensate for transmission losses, the XUV flux and thereby the total number of emitted electrons cannot be arbitrarily increased. Here, space-charge effects have to be considered, as discussed in the following section. Thus, for high spatial selectivity and a large DoF without significant transmission losses, the size of the XUV spot is an important parameter to consider. 

\section{Quantitative comparison of the MM and the HA} \label{sec:comparison}
\subsection{XUV-induced space charge} \label{sec:probe_spacecharge}

\begin{figure*}[t!]
\centering
\includegraphics[width=1\textwidth]{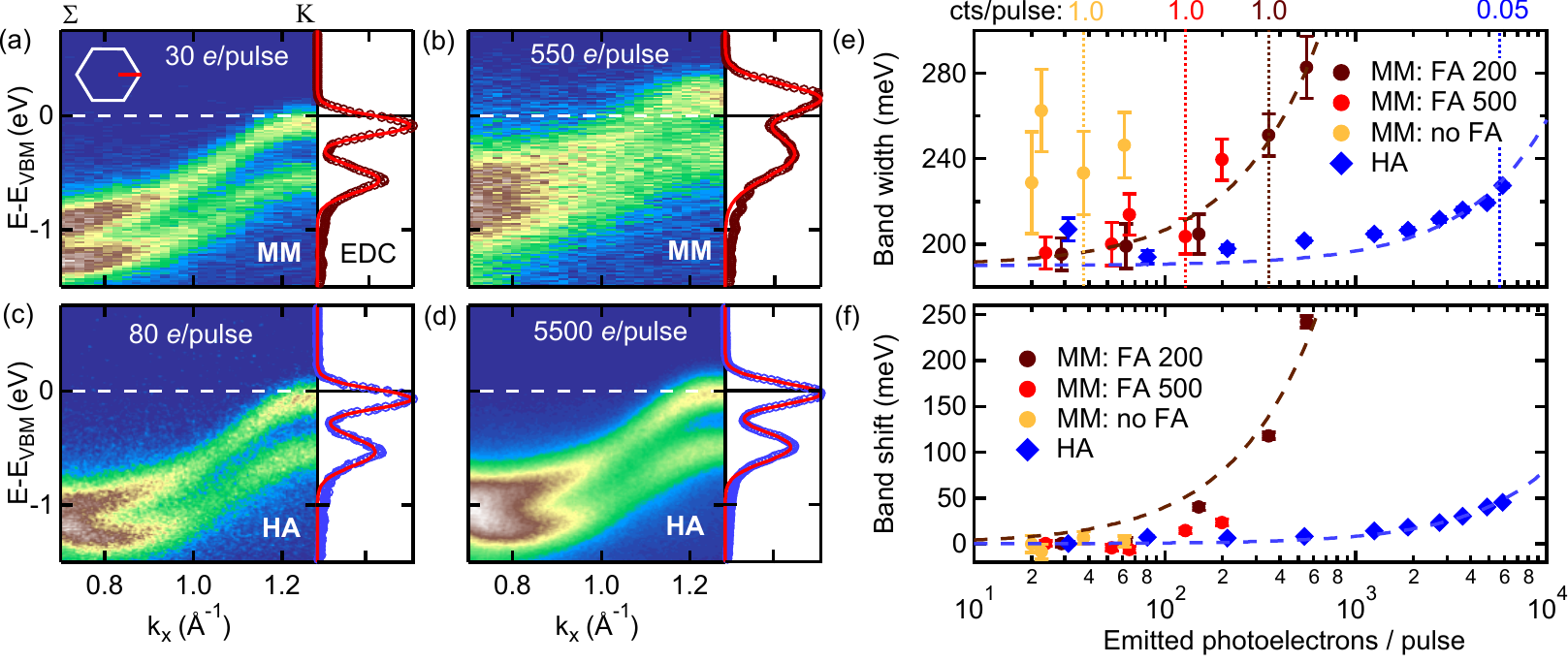}
\caption{(a,b) False-color plots of MM cuts ($d_{\mathrm{FA}}=200$\,\textmu m) and (c,d) HA measurements along the $\Sigma$-K direction at selected photoelectron emission rates. The EDCs at the K point (0.09\,\AA$^{-1}$ integration window) are shown with a fit by two Gaussians (red curve). (e) FWHM and (f) energy shift extracted from the fit to the upper band at the K point as a function of emitted photoelectrons per pulse (obtained from the sample photocurrent). The MM bandwidth values without an aperture deviate from the other curves, as the spectra are already significantly blurred due to the low DoF. Linear fits (dashed lines) serve as guides to the eye. Selected photoelectron detection rates are indicated by vertical lines. The effective electron detection rate of the HA is orders of magnitude below the MM, since a drastically smaller energy-momentum window is covered in a single measurement.} \label{fig:spacecharge_XUV}
\end{figure*}

A fundamental limitation of photoemission with ultrashort light pulses is space charge. The Coulomb repulsion within a dense photoelectron cloud can modify the electrons' angular and energy distribution, and can significantly deteriorate momentum and energy resolution. Space charge and its dependence on source parameters, such as pulse duration, flux, and spot size, have already been studied extensively\cite{zhou2005,passlack2006, hellmann2009, graf2010vacuum, rotenberg2014microarpes, plotzing2016spin}. Here, we compare the space-charge effects for both detection schemes using the energy shift and broadening of the energy dispersion curve (EDC) of the spin-orbit split VBs at the K point of WSe$_2$, see Fig.\,\ref{fig:spacecharge_XUV}. In the regime of few emitted photoelectrons per pulse, the band structure measurements of both detectors are in excellent agreement, see panels (a) and (c). When increasing the XUV source flux (and thereby the density within the photoelectron cloud), the MM spectrum rapidly shifts towards higher energies and becomes drastically broadened, while the HA spectrum is only weakly affected, see panels (b) and (d). 

For the MM, detectable energy distortions (shift and additional broadening $\gtrsim 10$\,meV) arise above $\sim$\,100 emitted electrons per pulse, roughly one order of magnitude before distortions appear in HA measurements, see Figs.\,\ref{fig:spacecharge_XUV}(e-f) and the discussion below. While the transmission and thereby the effective count rate decrease for a smaller field aperture size, we find that space charge is rather independent of the apertures. This demonstrates that its major contribution stems from the Coulomb interaction of photoelectrons on their trajectories prior to the Gaussian image plane, in agreement with simulation results for the case of hard X-ray ARPES\cite{schonhense2018}. In other words, to employ the MM at a reasonable resolution, the source flux and the resulting number of emitted photoelectrons per pulse have to be chosen carefully. For instance, when using the aperture $d_{\mathrm{FA}}$\,=\,$200$\,\textmu m (allowing for a large DoF), $\sim\, $350\,emitted photoelectrons per pulse are required to reach the instrumental limit of a single event per pulse due to transmission losses and an imperfect detection efficiency. However, in this regime, the spectrum is already significantly shifted and broadened.

\begin{figure}[t!]
\centering
\includegraphics[width=1\columnwidth]{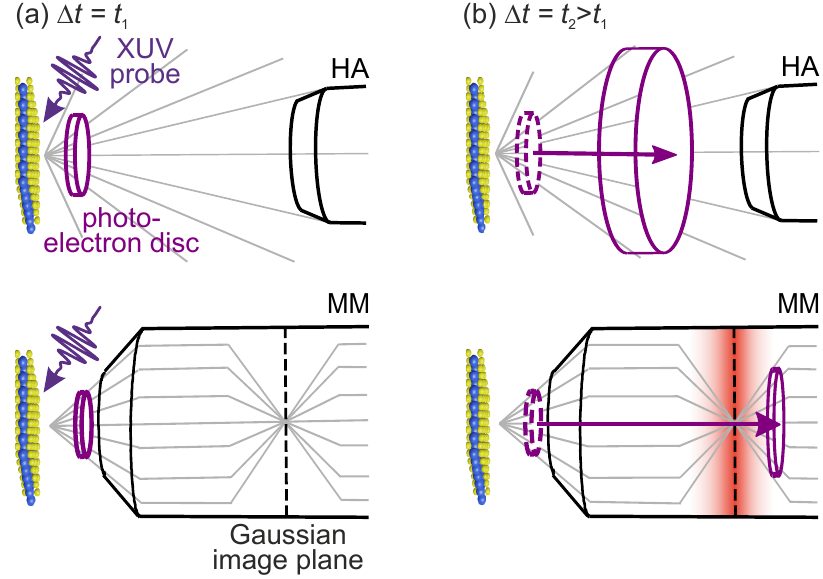}
\caption{(a) Schematic of the emitted photoelectron disc shortly after arrival of the XUV pulse for the HA (top) and MM (bottom). (b) Photoelectron disc at a later time. In the HA, the electron disc spreads over the complete 2$\pi$ solid angle and broadens along the direction of propagation. In the MM, all photoelectrons are guided into the lens column. Due to the high acceleration voltage, the relative energy difference between primary and secondary electrons is small, and the photoelectrons remain confined to a thin, dense disc. Focal planes (indicated in red) further increase the electron-electron interaction.}
\label{fig:spacecharge_schematic}
\end{figure}

These observations demonstrate that space-charge effects can be a major limitation of the MM compared to HAs when using fs pulses, as illustrated in Fig.\,\ref{fig:spacecharge_schematic}. In the field-free region in front of the HA, the emitted photoelectron disc spreads over the complete 2$\pi$ solid angle, while it simultaneously broadens along the direction of propagation due to the high relative energy difference between primary (fast) and inelastically scattered, secondary (slow) electrons. In contrast, the high extractor field of the MM guides the entire electron cloud into the lens column, leading to increased electron densities, and accelerates it to few keV. Therefore, before discrimination in the ToF-tube, the primary and secondary electrons propagate at comparable velocities and spread only marginally along the direction of propagation. As a result, the effective interaction travel length and interaction time between fast electrons of the primary spectrum and secondary electrons is significantly higher in the MM. Refocusing of the photoelectron disc at several focal planes of the lens column further increases these space-charge effects. Since the secondary electrons travel close to the optical axis, the primary electron spectrum features a Lorentzian profile of iso-energy surfaces, with space-charge distortions most pronounced at the $\Gamma$ point. Note that these deterministic energy shifts can be compensated by numerical correction\cite{schonhense2018}, in contrast to the space-charge-induced energy broadening.

For illustration, we estimate the involved time and length scales of the spread of the photoelectron cloud along its trajectory for both instruments. In the HA, it takes $\approx$90\,ps to separate electrons of highest kinetic energy (corresponding to the VB maximum, $E_{\mathrm{kin}}\approx17$\,eV) from the secondary electron tail (exemplary energy $E_{\mathrm{kin}}\approx5$\,eV) by 100\,\textmu m, during which the fast electrons have traveled 220\,\textmu m. In the MM (at an approximate average potential within the initial lens elements of $2$\,kV after acceleration), it takes $\approx$1.3\,ns to achieve the same distance spread between fast and slow electrons, during which the primary electrons have traveled several centimeters, reaching already into the lens column. Note that when using the MM in the regime of soft/hard X-ray PES, the relative velocity difference of primary and secondary electrons is increased, which reduces the space-charge interaction between the two electronic species. Detailed simulations of the space-charge effects in momentum microscopy can be found elsewhere\cite{schonhense2018}.

In summary, in our experimental configuration (XUV spot diameter of 80\,\textmu m), we apply a small FA to reach a sufficient DoF in typical band mapping experiments. While the resulting transmission losses can be partially compensated by increasing the number of total emitted photoelectrons per pulse, space-charge effects ultimately constrain the operating conditions to a regime significantly below the detector saturation. A reduction of the spot diameter below 25\,\textmu m would allow to omit field apertures in most experiments. Due to the increased transmission, only few emitted photoelectrons per pulse are required to reach the limit of detector saturation, and space-charge effects would be negligible with respect to the typical energy resolution in time-resolved experiments. However, most current trARPES setups work at spot diameters in the range of 80\,\textmu m to few hundred \textmu m\cite{corder2018, buss2019, cucini2019, puppin2019, keunecke2020}, as reducing the focus size below few 10\,\textmu m in the XUV regime is difficult to achieve with conventional beam line layouts.

\subsection{Count rates}
\label{sec:countrates}

Next, we discuss the total count rate of both instruments achievable under these space-charge restrictions. For the MM, the detected counts per pulse $cts_\mathrm{MM}$ are given by the product of the total emitted counts per pulse, the transmission of the FA, and the quantum efficiency of the DLD/MCP stack

\begin{equation}
  cts_\mathrm{MM}=e_\mathrm{tot} \cdot T_\mathrm{FA} \cdot QE_{\mathrm{MM}}\,.
\end{equation}

We estimate $QE_{\mathrm{MM}}\approx5$\,\% from measurements without an FA ($T_\mathrm{FA}\approx1$). Combined with the transmission losses at the FA in typical experiments ($T_{\mathrm{200 \mu m}}\approx6$\,\%), roughly 0.3\,\% of the total emitted photoelectrons are detected.

For the HA, the detected counts per pulse are given by
\begin{equation}
  cts_\mathrm{HA}=e_\mathrm{tot} \cdot f \cdot QE_{\mathrm{HA}}\,,
\end{equation}
with the fraction of the electron distribution sampled in a single HA measurement $f$ and the quantum efficiency of the MCP $QE_{\mathrm{HA}}\approx10$\,\%. For our settings of a slit width of $0.5\,\mathrm{mm}$ and a pass energy of $30\,\mathrm{eV}$ centered on the upper valence band region, we estimate $f\approx 0.03$\,\%. Therefore, 0.003\,\% of the total emitted photoelectrons are detected, which is roughly two orders of magnitude below the MM. Note that the relatively low quantum efficiency of the detectors in both instruments could be related to detector aging and the comparably low impact energy of the photoelectrons\cite{Goruganthu1984}.

In our measurement configuration, the space-charge limit of $e_\mathrm{tot} \approx 100$ emitted photoelectrons per pulse restricts the MM count rate to $cts_\mathrm{MM} \approx 0.3$\,cts/pulse. However, a substantial portion of these electrons originates from the secondary electron tail and deep-lying VBs. Focusing only on the topmost VB region from the VB maximum to 1.5\,eV below, which is typically of most interest in time-resolved studies, yields 0.006\,cts/pulse, or 3000\,cts/s at a repetition rate of 500\,kHz. In contrast, when using the HA, the XUV flux can be increased by approximately an order of magnitude to $e_\mathrm{tot} \approx1000$ photoelectrons per pulse before critical space-charge effects emerge. For a typical cut, such as shown in Fig.\,\ref{fig:setup}(e), we detect approximately 15000\,\,cts/s. In comparison, when extracting a comparable cut from the MM dataset, the count rate is roughly 40\,cts/s, which is a factor of $\approx350$ below the rate of the HA, resulting from the reduced XUV source flux ($\sim10\times$), transmission losses at the aperture ($\sim17\times$) and a lower detection efficiency ($\sim2\times$). Nevertheless, when the total photoelectron distribution is of interest, the MM outperforms the HA by a factor of $cts_\mathrm{MM} / cts_\mathrm{HA}\approx10$.

In an optimal scenario (small XUV spot, $T_{\mathrm{FA}}\approx1$, $QE_{\mathrm{MM}}\approx1$), only few emitted photoelectrons per pulse are required, and the count rate of the MM $cts_\mathrm{MM}$ is only limited by the detector saturation of $\sim$1\,cts/pulse. However, compared to the experimental scenario discussed above (0.3\,cts/pulse), this increases the total count rate only by a factor of $\approx 3$. Thus, also under optimized conditions our conclusions still hold true. Another approach to improve the MM count rate is by increasing the repetition rate of the laser system. However, in pump-probe experiments, re-equilibration of the sample within the laser's duty cycle has to be considered, which, at multi-MHz repetition rates, critically limits the applicable excitation fluences. While, in the regime of very weak excitation, repetition rates of several 10 MHz allow to mitigate space-charge effects and to substantially reduce acquisition times\cite{corder2018, mills2019}, the ToF of slow electrons further limits the maximum applicable repetition rate (typically $< 10$\,MHz)\cite{tusche2016multi}.

\subsection{Experimental scenarios}
\label{sec:use_cases}

\begin{figure}[t!]
\centering
\includegraphics[width=1\columnwidth]{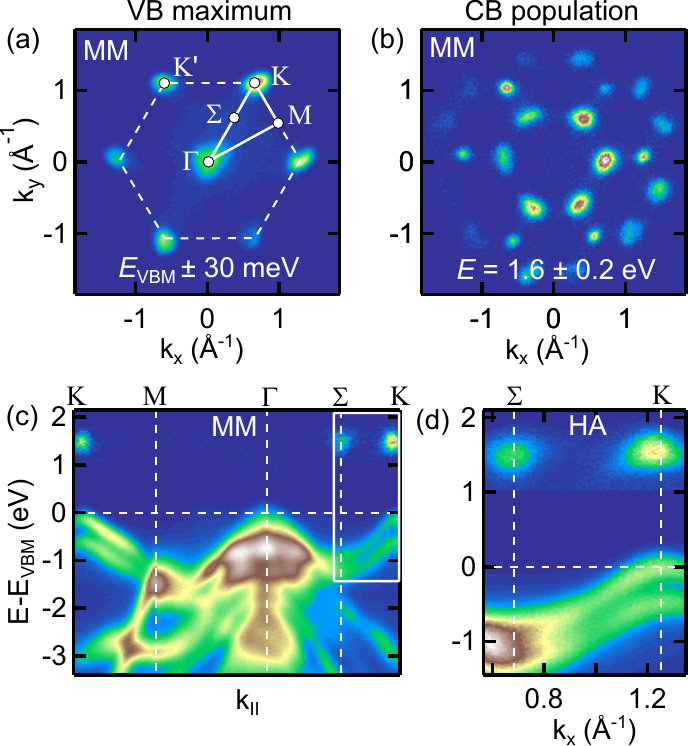}
\caption{MM iso-energy contours of (a) the VB maximum and (b) the transiently excited CB population. Asymmetries between equivalent points result from orbital interference effects in photoemission\cite{beaulieu2020WSe2}. Minor symmetry distortions were corrected using symmetry-guided registration\cite{xian2019symmetry}. (c) Extracted energy-momentum cuts along the high-symmetry directions. (d) Data acquired with the hemispherical analyzer, corresponding to the energy-momentum window indicated by the white box in panel (c). The intensity of the CBs is enhanced by a factor of 75 in both datasets to achieve a comparable intensity of the CB signal and the topmost VBs at K ($t=0\pm 50$\,fs, $h\nu=1.55$\,eV, $F_{\mathrm{abs}}=150$\,\textmu J/ cm$^{2}$). For comparability, the momentum integration orthogonal to the plotted direction of the MM cuts in (c) is matched to the HA measurements.}
\label{fig:800nm_mapping}
\end{figure}

Next, we discuss common trARPES scenarios to highlight the advantages of each instrument and the benefit of combining both detectors in a single setup. As a first use case we show the (excited-state) band mapping of bulk WSe$_2$ upon excitation with near-infrared optical pulses ($\lambda_{\mathrm{pump}}=800$\,nm). Using the MM, we acquire the quasiparticle dispersion across the full photoemission horizon in a single measurement at a fixed sample geometry. We gain access to the band structure of the first projected BZ up to 1.55\,eV above the VB, see the transient occupation of the CB at the K and $\Sigma$ points in Figs.\,\ref{fig:800nm_mapping}(b,c). For static 3D band mapping using the MM, typically $10^7-10^8$ total events are required, as a large portion of the intensity originates from secondary electrons, achievable in $\sim1-10$\,minutes at a typical count rate of $\approx 1.5\cdot 10^5$\,cts/s (Fig.\,\ref{fig:800nm_mapping}(a)). In order to accurately resolve the much weaker signal of excited states, typically $\sim10^9$\,events are detected within $\sim 2$\,hours, producing data as shown in Fig.\,\ref{fig:800nm_mapping}(b,c). For comparison, the energy-momentum window covered in a single HA measurement along the $\Sigma$-K direction is shown in panel (d), recorded within $\sim10$\,minutes. Mapping the full irreducible part of the BZ with the HA (by sample rotation or by using a deflector arrangement) requires at a comparable momentum resolution $\sim60$ sequential scans. This procedure is further complicated by the fact that high emission angles are difficult to access and spectra have to be merged and mapped from angle to momentum space. In addition, light polarization, fluence and photoemission matrix elements might change during such a mapping procedure using a sample manipulator. Thus, to get an overview of the full (excited-state) dispersion relation, band mapping with the MM is highly advantageous.

\begin{figure}[t!]
\centering
\includegraphics[width=1\columnwidth]{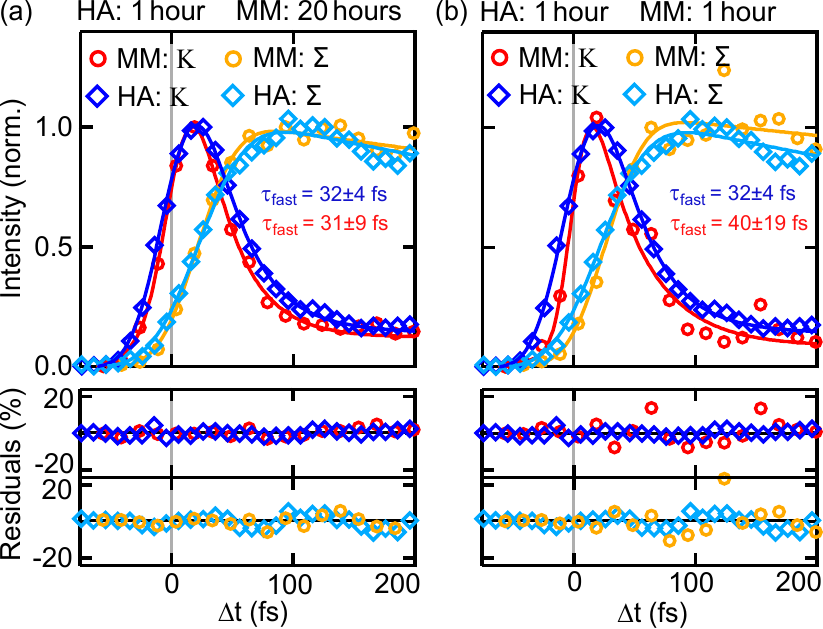}
\caption{Time traces of the integrated excited-state signal at the K and $\Sigma$ valleys for acquisition times of (a) 1\,hour (HA) and 20\,hours (MM), and (b) an equal acquisition time of 1\,hour for both instruments. The excited-state signal of the MM data is extracted from an energy-momentum plane corresponding to the HA measurement. The time traces are fitted with a single-exponential ($\Sigma$) and double-exponential (K) decay curve convolved with a Gaussian, respectively. The fit results are shown in solid curves, along with the time constants (standard deviation as uncertainty) of the fast decay component of the transient population at K. While the residuals in panel (a) show similar levels of noise for both instruments, the MM time traces in (b) feature substantial scatter.}
\label{fig:800nm_timetrace}
\end{figure}

Another typical use case of trARPES is the investigation of the transient carrier relaxation dynamics along certain pathways in momentum space. In bulk WSe$_2$, electrons are initially excited into the conduction band (CB) at the K valley, followed by a relaxation into the global CB minimum at the $\Sigma$ point. Note that while we study a bulk sample, the dominant fraction of the photoemission intensity originates from the topmost WSe$_2$ layer due to the limited photoelectron escape depth in the XUV regime.  A detailed discussion of the relaxation dynamics in bulk and monolayer WSe$_2$ can be found elsewhere\cite{bertoni2016, puppin2017time, madeo2020}. As such relaxation dynamics are often highly localized in momentum space, information on selective regions in momentum space is sufficient to study the temporal evolution. Measuring such dynamics with the MM results in a 4D data set (3D + time) of the full energy-, momentum- and time-dependent band structure, which requires $\sim10^{10}$ events and an acquisition time of 20\,hours or more, depending on the sample characteristics, the required statistics and the pump-probe delay range. In contrast, using the HA, only the relevant energy-momentum region is recorded, and we can utilize the higher photon flux and larger transmission within this window, yielding an acquisition time in the range of 1-2\,hours for a time trace. To illustrate these differences, Fig.~\ref{fig:800nm_timetrace}(a) shows the time traces of the conduction band population at the K and $\Sigma$ points for both spectrometers, measured for 1 hour (HA) and 20 hours (MM), respectively. Both data sets show similar statistics and scatter, as visible from the residuals of the exponential fits. In contrast, comparing the data for similar acquisition times (Fig.~\ref{fig:800nm_timetrace}(b)) shows much larger scatter in the MM traces due to the lower number of acquired events. This is also represented in the accuracy of the fit parameters. Even if we sum the symmetry-equivalent locations in the Brillouin zone that the MM data cover, the HA still permits much faster data acquisition of a limited energy-momentum region. This allows for a time-dependent systematic variation of external parameters (e.g. temperature, pump fluence, etc.) -- challenging with the MM. However, if the electron dynamics over an extended momentum-space region are of interest\cite{Rettig2016} or comparing different momentum points not simultaneously accessible within the angular range of the HA is required\cite{bertoni2016}, the MM is clearly advantageous.

\begin{figure*}[t!]
\centering
\includegraphics[width=1\textwidth]{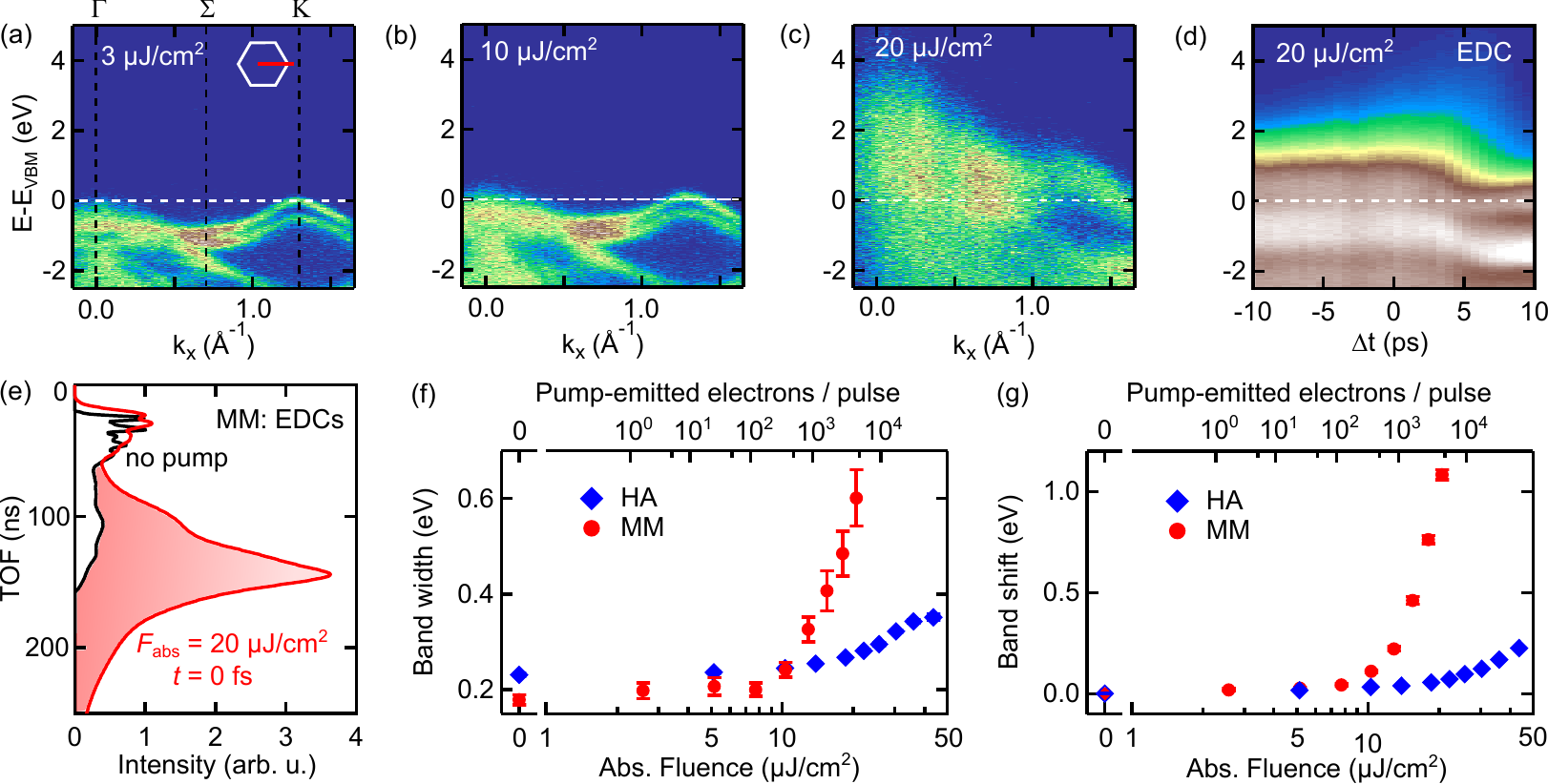}
\caption{(a-c) False-color plots of 2D MM cuts along the momentum direction indicated by the red line in panel (a) for various pump fluences ($h\nu=3.1$\,eV, $t=0$\,fs). (d) Total momentum-integrated EDC as a function of time delay. (e) Total intensity versus ToF at equilibrium (black) and with optical pump at $t=0$\,fs (red). (f,g) Fit results (analogous to Fig.\,\ref{fig:spacecharge_XUV}) of width and shift of the VB at the K point as a function of absorbed fluence ($t\approx0$\,fs, $d_{\mathrm{FA}}=200$\,\textmu m) and of pump-laser-emitted electrons per pulse. The sharp onset of the space-charge effects in the MM measurements demonstrates the high nonlinearity of the pump-pulse-induced photoemission.}
\label{fig:spacecharge_pump}
\end{figure*}

\subsection{Optical pump-induced space charge} \label{sec:pump_spacecharge}

A further critical aspect in trARPES are the space-charge effects induced by electrons emitted by the pump pulses. Multi-photon photoemission and emission at surface inhomogeneities can generate a significant number of low-energy electrons. Depending on the pump-probe time delay, this can lead to complex interactions with the probe-pulse-induced electron cloud\cite{al2015, oloff2016pump, kutnyakhov2020}. While this phenomenon plays a secondary role when exciting WSe$_2$ at $h\nu=1.55$\,eV, it becomes increasingly important as the photon energy of the pump pulses approaches the material's work function, since the order of the nonlinearity needed for multiphoton ionization decreases. In the following, we systematically study the pump-induced space-charge effects at $h\nu=3.1$\,eV, utilizing the metrics introduced in Sec.\,\ref{sec:probe_spacecharge}, i.e., the energy shift and broadening of the VB at the K point.

Already at moderate excitation densities ($F_{\mathrm{abs}}=20$\,\textmu J/cm$^2$), the MM spectra exhibit a severe non-uniform broadening and shift most pronounced at the $\Gamma$ point, see Fig.\,\ref{fig:spacecharge_pump}. In this fluence regime, the low-energy electrons released by the pump pulses greatly outnumber probe-pulse-induced photoelectrons, shown in panel (e). The pump-pulse-induced space-charge effects strongly depend on delay\cite{kutnyakhov2020} and extend over several ps around the temporal pump-probe overlap, see panel (d). Here, one has to carefully distinguish the true temporal overlap from the space-charge maximum at positive delays. Space-charge interaction is particularly critical at positive delays (pump pulse precedes the XUV probe), since the fast probe photoelectron cloud traverses through the cloud of slow, pump-pulse-emitted electrons on its path to the detector. In the MM, the relative difference between the velocities of the two electron species is minute due to the high acceleration field of the extractor, similar to the interaction between the primary and secondary electrons within the probe-pulse electron cloud discussed in Sec.\,\ref{sec:probe_spacecharge}. As a result, the critical interaction region extends far into the lens column. Moreover, also the low-energy electrons emitted by the pump pulses travel along the optical axis, which enhances the energy shift and broadening at the $\Gamma$ point. In contrast, in the field-free region between the sample and the HA, the relative speeds of both electron clouds differ strongly, so the interaction region is limited to a small volume close to the sample. Thus, pump-pulse-induced space-charge effects completely blur the band dispersion already at several 100 pump-emitted electrons per pulse when using the MM, while distinct bands can be discerned with the HA above 10000 electrons per pulse, as shown in Figs.\,\ref{fig:spacecharge_pump}(f-g), in agreement with XUV-induced space-charge discussed in Sec.\,\ref{sec:probe_spacecharge}.

\begin{figure}[t!]
\centering
\includegraphics[width=1\columnwidth]{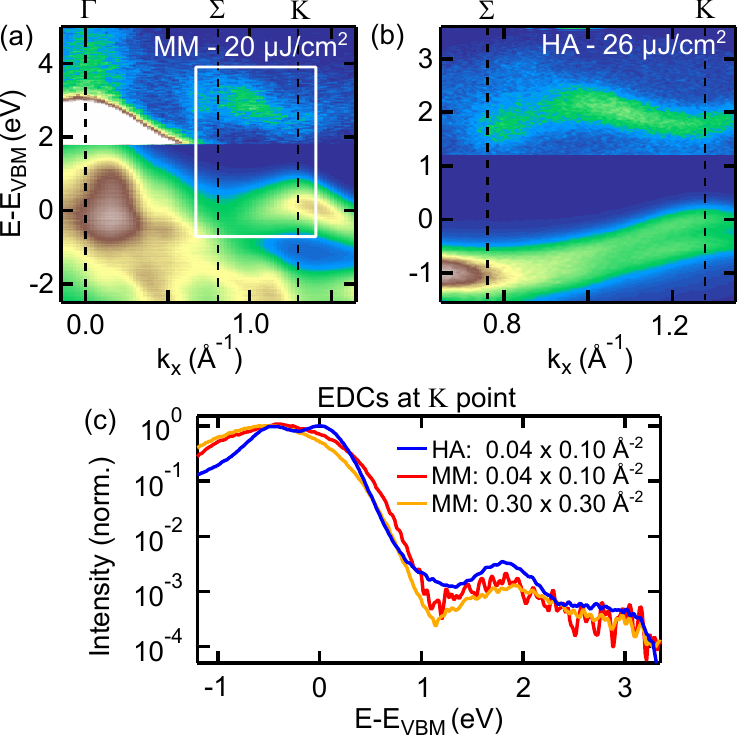}
\caption{Excited-state band mapping along the $\Gamma-\Sigma-\mathrm{K}$ direction ($t=0$\,fs, $h\nu=3.1$\,eV) using (a) the MM (20\,hours acquisition) and (b) the HA (4\,hours). The CB intensity is enhanced by a factor of 250 to improve visibility. The white box in panel (a) indicates the area covered by the HA measurement. (c) EDCs extracted at the K point. Despite similar excitation densities, the excited-state signal at 1.8\,eV of the MM traces shows significant scatter, both for a small (red) and an extended momentum-integration (orange curve) around the K point.}
\label{fig:400nm_mapping}
\end{figure}

Ultimately, this significantly limits the experimental flexibility of the MM with regard to excitation wavelengths approaching the sample work function, and strongly restricts the applicable excitation fluences. For our test case of $h\nu=3.1$\,eV, two-photon processes dominate the pump-induced photoemission from WSe$_2$. Here, pump-induced space charge strongly shifts and distorts the spectra near the $\Gamma$ point, and at the same time heavily blurs the dispersion at K, which makes it difficult to discern the excited-state signal at $10^{-3}$ of the level of the VB. In contrast, at comparable excitation densities, the HA delivers a sharp band dispersion, a well-resolved spin-orbit splitting of the VB, and a clear excited-state signal within reasonable integration times, illustrated in Fig.\,\ref{fig:400nm_mapping}. Also, the HA permits significantly increased excitation fluences creating a larger excited-state population without considerable distortions and allows even higher excitation photon energies providing a larger window into the conduction band dispersion.

\section{Discussion} \label{sec:discussion}

Our case studies show that for trARPES studies, despite parallel detection of the full energy and momentum range, the ToF-MM in its current state does not replace, but rather complement the HA. Moreover, the combination of the two complementary detection schemes in a single setup allows us to address a broad variety of scientific questions. To illustrate the complementary role of both instruments, let us consider the scenario of studying a novel material. For an initial characterization, the MM is best suited, as it permits an efficient mapping of the full band structure and gives an overview of all relevant carrier relaxation pathways within the entire projected BZ. After identifying central energy-momentum regions with the MM, the HA can be used to quickly analyze the dynamics within specific energy-momentum regions at high momentum resolution, and to systematically explore the experimental parameter space (e.g. fluence and temperature dependence) in a time-resolved fashion. Moreover, the HA can provide access to experimental parameter ranges, e.g., excitation wavelengths, fluences and polarizations, that are not feasible using the MM due to space-charge restrictions or the experimental geometry (grazing-incidence illumination). 

A complementary advantage of the MM is the possibility to measure samples that are susceptible to XUV beam damage, as only a very limited XUV exposure is required due to the efficient simultaneous detection of the full photoemission horizon. In addition, we note here also a few additional experimental difficulties connected with the MM. Firstly, flat sample surfaces are needed to prevent field emission resulting from the high extractor voltage. Secondly, a flat and isotropic sample holder is required to prevent distortions of the extractor fields. Thirdly, acquisition with the MM requires processing and storage of large data sets ($\sim$\,100\,GB for a typical data set of $10^{10}$ events), and involves complex data binning and analysis procedures\cite{xian2019open, xian2019symmetry}.

We demonstrated that space-charge effects and the detector saturation critically limit the experimental count rates of the MM in particular for time-resolved studies. Future developments of high-pass filtering of secondary and pump-emitted photoelectrons close to the sample are expected to mitigate these limitations. Combining such filtering techniques and DLDs with improved multi-hit capabilities would be necessary to exploit all benefits of the MM also under experimental conditions comparable to the HA.

Determination of the complete (time-resolved) electronic band structure dynamics with the MM bears an enormous potential. Most directly, it allows to track complex momentum- and energy-dependent scattering phenomena, shines light on quasiparticle lifetimes\cite{haag2019}, and permits benchmark comparison to band structure theory\cite{xian2020}. As the MM measurements are performed at a fixed sample geometry, it allows to investigate higher-order modulation effects of the photoemission intensity, such as orbital interference\cite{beaulieu2020WSe2}. In addition to comprehensive band structure mapping, the MM bears further conceptually new measurement configurations. The use of apertures in the real-space image plane permits spatial selectivity of band structure measurements down to the micrometer scale. Furthermore, the use of apertures in the reciprocal image plane allows to extract the real-space photoelectron distribution at high momentum-selectivity via PEEM. This novel technique allows to study spatial inhomogeneities that involve subtle momentum-variations, such as the formation of domain boundaries of symmetry-broken states, the impact of defects on ordering phenomena, and the spatial distribution of intertwined complex phases after photoexcitation\cite{fradkin2015colloquium, wandel2020light}.

\section{Conclusion} \label{sec:conclusion}
We have demonstrated a dual-detector XUV time-resolved ARPES setup, and benchmarked the characteristics of a time-of-flight electron momentum microscope and a hemispherical analyzer, using metrics such as depth of focus, pump- and probe-pulse-induced space-charge effects, and experimental acquisition times. The unique combination of analyzers enables a full view of the band structure dynamics across the entire photoemission horizon using the momentum microscope and a rapid data acquisition across a limited energy-momentum region at high momentum resolution using the hemispherical analyzer. Furthermore, the possibility to achieve high spatial selectivity and the option of mapping the (time-dependent) real space photoelectron distribution of confined spectral features via momentum-resolved photoemission electron microscopy allow for entirely new perspectives.

\section*{Supplementary material}
See the supplementary material for a video of the temporal evolution of the excited-state signal in WSe$_2$ acquired with the MM (iso-energy contour at $1.6 \pm 0.2$\,eV) and the HA (cut along the $\Sigma$-K direction, CB signal enhanced).

\section*{Data availability statement}
The data that support the findings of this study are publicly available\cite{maklar_julian_2020_4067968}.

\section*{Acknowledgements}
We thank S. Kubala, M. Krenz, D. Bauer, R. Franke, J. Malter, M. Wietstruk (SPECS GmbH), A. Oelsner, M. Kallmayer, and M. Ellguth (SurfaceConcept GmbH, Mainz) for technical support. We also thank G. Schönhense (University of Mainz) for enlightening discussions. This work was funded by the Max Planck Society, the European Research Council (ERC) under the European Union's Horizon 2020 research and innovation program (Grant No. ERC-2015-CoG-682843), the German Research Foundation (DFG) within the Emmy Noether program (Grant No. RE 3977/1), the DFG research unit FOR 1700, and through the SFB/TRR 227 "Ultrafast Spin Dynamics" (projects A09 and B07). S.B. acknowledges financial support from the NSERC-Banting Postdoctoral Fellowships Program.

\subsection*{Corresponding authors}
maklar@fhi-berlin.mpg.de\\ rettig@fhi-berlin.mpg.de

\newpage
\onecolumn
\printbibliography{}

\end{document}